\documentclass[runningheads]{llncs}
\usepackage[T1]{fontenc}

\usepackage{amsmath,amssymb,amsfonts}
\usepackage{algorithmic}
\usepackage{graphicx}
\usepackage{textcomp}
\usepackage{multirow}
\usepackage{subcaption}

\usepackage{xcolor} 
\definecolor{darkgreen}{rgb}{0,0.5,0} 
\definecolor{darkyellow}{rgb}{0.9, 0.7, 0.0}   

\begin{document}

\title{Word Synchronization Challenge: A Benchmark for Word Association Responses for Large Language Models}
\titlerunning{WSC: A Benchmark for Word Association Responses for LLMs}
%
\author{Tanguy Cazalets\inst{1,2}\orcidID{0000-0001-5614-5573} \and
Joni Dambre\inst{1,2}\orcidID{0000-0002-9373-1210}}
\authorrunning{Tanguy Cazalets et al.}
\institute{Ghent University, Gent, Belgium \and
Airo lab, Gent, Belgium}
\maketitle              
\begin{abstract}
This paper introduces the Word Synchronization Challenge, a novel benchmark to evaluate large language models (LLMs) in Human-Computer Interaction (HCI). This benchmark utilizes a dynamic game-like framework to test LLMs’ ability to mimic human cognitive processes through word associations. By simulating complex human interactions, it assesses how LLMs interpret and align with human thought patterns during conversational exchanges, essential for effective social partnerships in HCI. Initial findings highlight the influence of model sophistication on performance, offering insights into the models’ capabilities to engage in meaningful social interactions and adapt behaviors in human-like manners. This research advances understanding of LLMs’ potential to replicate or diverge from human cognitive functions, paving the way for more nuanced and empathetic human-machine collaborations.
\end{abstract}

\section{Introduction}

In the evolving field of Human-Computer Interaction (HCI), seamless communication with humans is essential for creating autonomous systems that are both user-friendly and effective \cite{obaigbena2024ai}. A critical aspect of this communication is chatbots’ ability to interpret and respond to human language, a task increasingly powered by large language models (LLMs) \cite{kim2024understanding}. These models have revolutionized the way machines handle language, enabling them to process complex patterns, word associations, and semantic structures \cite{hadi2024large}. However, as LLMs become more integral to HCI, evaluating their performance in understanding human language and generating human-like responses has become a key focus of researchers \cite{zhang2023large}.

Assessing the effectiveness of LLMs goes beyond just measuring their linguistic accuracy. It involves evaluating how well they capture word associations and reflect the cognitive processes humans use in communication \cite{shankar2024validates}. By analyzing these models’ ability to mimic human thought and emotional understanding, researchers can ensure that LLMs can engage in meaningful conversations, adapt to user behaviors, and respond in ways that feel natural and considerate, enhancing their role as social partners in human environments \cite{kim2024understanding}.

Word associations, a fundamental aspect of both linguistic research and psychological studies \cite{clark1970word}, illuminate how concepts are interconnected within human cognition. They provide essential insights that can be used to program chatbots that mimic human thought processes and emotional responses. This capability is vital for chatbots to engage in meaningful dialogues and to act as social partners, adapting their behaviors in ways that are perceived as natural and considerate by human users.

We introduce the Word Synchronization Challenge, a novel benchmark designed to assess LLMs' ability to capture and utilize word associations dynamically. By simulating cooperative word association in a game-like framework, we test models' capacities to align with human thought patterns and assess their implicit theory of mind capabilities \cite{Weerd2015}  \cite{Street2024}.

After providing the necessary background in Section \ref{background}, we detail the benchmark, its various settings, and its implementations in Section \ref{methods}. We then present an analysis of the first results collected in Section \ref{results}, and discuss its broader implications for future HCI advancements and future directions in Section \ref{discussion}. 

\section{Background} \label{background}

\subsection{Word Associations in Human Cognition}
Word associations, as a reflection of semantic memory, serve as a fundamental element in understanding how humans generate and process language. In traditional psychological studies, word associations are used to uncover the underlying connections in human memory and the associative networks that guide our thinking and language use \cite{clark1970word}. 

Recent advances in word association research provide valuable insights into lexical access and semantic networks, which are crucial for improving LLM performance \cite{pranoto2019organization}. In a large-scale study involving over 12,000 cue words and responses from 70,000 participants, researchers constructed a comprehensive semantic network using multiple-response free association tasks. This approach revealed that multiple responses produce more diverse and accurate predictions of lexical access and semantic relatedness, surpassing single-response methods. These findings are particularly relevant for LLMs, which benefit from such variability to infer semantic relationships, though they still face challenges in handling the nuanced associations and contextual shifts that humans navigate naturally.\cite{DeDeyne2012}

\subsection{Large Language Models and Word Association Performance}

The advent of LLMs like GPT (Generative Pre-trained Transformer) and BERT (Bidirectional Encoder Representations from Transformers) has revolutionised the field of HCI \cite{zhang2023large}. These models, trained on extensive corpora, develop an ability to predict word associations that often mimic human thought processes. However, their performance varies based on the depth of training and the diversity of the training data. 

Studies have shown that while LLMs are proficient at generating syntactically correct responses\cite{PerezMayos2021} \cite{Janssens2024}, their ability to mirror true human-like associative thinking remains a challenge. The Word Synchronization Challenge addresses this gap by evaluating LLMs not only on their linguistic output but on their capacity to dynamically align with human associative patterns during interactions. This evaluation reflects a shift from traditional static performance metrics towards dynamic, interaction-based assessments that are crucial for real-world applications of HCI.

Although the study of word associations is well-established in psychology and linguistics, offering crucial insights into human cognitive processes, remarkably few studies have explored these associations through the lens of LLMs. A recent comprehensive review of related work is provided by \cite{Abramski2024}. This gap in the research highlights the importance of our benchmark in contributing to a more nuanced understanding of LLMs' capacities, bridging the divide between human cognitive functions and artificial linguistic processing.

\subsection{Motivations}

The implications of word associations in AI extend beyond technical performance and touch upon ethical and cultural dimensions. Since LLMs learn from datasets that may contain biases, their associative outputs can inadvertently perpetuate stereotypes or reflect cultural biases. By closely examining these associations, researchers can identify and mitigate unwanted biases in AI models, ensuring that these technologies act in ways that are fair, unbiased, and aligned with societal values.

The Word Synchronization Challenge not only serves as a benchmark for technological advancement but also provides a foundation for a myriad of related tasks, including:

\begin{itemize}
\item \textbf{Social Interaction Evaluation:} The primary goal of this benchmark is to assess the LLMs’ capacity to predict and adapt within social interactions, using their response to word associations to gauge their ability to mirror human reasoning and adapt their behaviors accordingly. These interactions are not merely linguistic tasks but are integral to developing chatbots capable of nuanced social participation and empathy.
\item \textbf{Model Understanding:} It helps us understand how LLMs organize and relate information. By examining word associations, researchers can gain insights into the underlying mechanisms of these models, such as how they capture semantic relationships or encode context. \cite{Thawani2019} \cite{Yao2022}
\item \textbf{Comparison with Human Cognition:} Studying word associations in LLMs allows for comparisons with human cognitive processes. This can inform whether LLMs mimic human-like associative patterns or if they develop distinct mechanisms for linking concepts. \cite{Abramski2023} \cite{Vintar2024}
\item \textbf{Bias and Fairness:} Word associations can reveal biases encoded in the training data. For instance, if a model consistently associates certain professions or activities with a particular gender or ethnicity, this could indicate biases that need addressing to ensure the model’s fairness and reliability. \cite{Piermatteo2018} \cite{Kaneko2021} \cite{Abramski2023}
\item \textbf{Language and Culture:} Word associations can reflect cultural contexts and linguistic structures, offering a window into how language is shaped by and shapes societal norms and values. \cite{Piermatteo2018}
\end{itemize}

\section{Methods} 
\label{methods}

\subsection{Game Description}
The task entails a dyadic interaction where two participants, which can include the LLM, are involved in a repeated word production game.

\textbf{Goal :}
The primary objective of the game is for both participants to articulate the same word in a round.

\textbf{Rules :}
\begin{itemize}
    \item \textbf{Word Production}
    \begin{itemize}
        \item Each participant must produce a word.
        \item Words produced in previous rounds cannot be used again.
        \item In the first round, participants can choose any random word.
    \end{itemize}
    
    \item \textbf{Evaluation}
    \begin{itemize}
        \item After both participants have spoken their words, the words are compared.
        \item If both participants say the same word, the game ends and they have achieved the goal.
        \item If the words are different, the game proceeds to the next step.
    \end{itemize}
    
    \item \textbf{Reflection}
    \begin{itemize}
        \item Participants engage in a preparatory phase for the upcoming round.
        \item During this phase, each participant reflects on the words spoken in the previous round and tries to understand the thought pattern of the other participant. This understanding will guide their word choice for the next round.
    \end{itemize}
\end{itemize}

\begin{figure}[ht]
  \centering
  \includegraphics[width=0.35\linewidth]{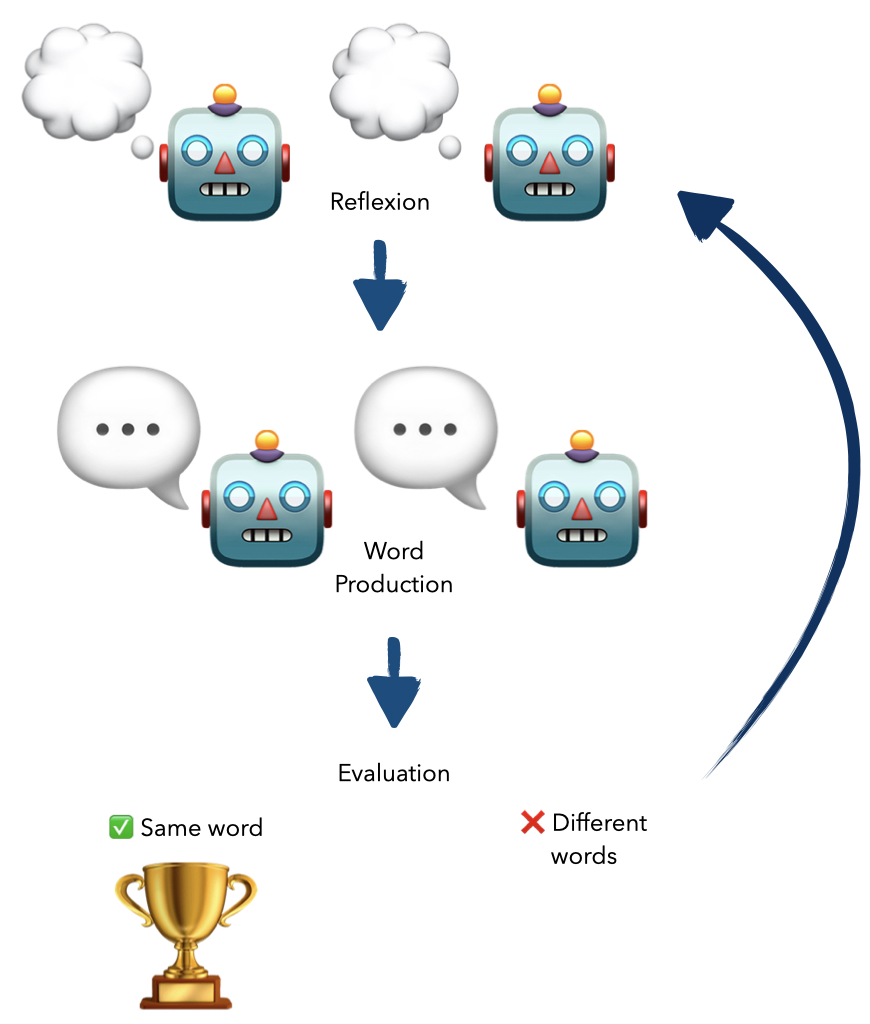}
  \caption{An illustration of the game rules}
  \label{fig:rules}
\end{figure}

\subsection{LLMs Word Associations Dataset Generation}

We developed a dynamic game simulation to generate a comprehensive dataset for evaluating model performance in word association synchronization. The dataset captures the full history of words exchanged between two LLMs.

For this study, we employed various LLMs from OpenAI "gpt-4o-mini", "gpt-3.5-turbo-0125", "gpt-4-turbo", testing same model and different models interactions. Throughout each game, the complete sequence of word associations and the final outcome (either \textit{won}, \textit{lost due to repetition}, or \textit{lost due to non-convergence}) was recorded. This dataset serves as the basis for evaluating how effectively different models synchronize in word association tasks, providing valuable insights into their behavioral patterns and overall performance.

The characteristics of each model combination are summarized in Table \ref{tab:model_properties}

\begin{table}[ht]
\centering
\caption{Properties of GPT Models}
\begin{tabular}{|l|l|l|l|}
\hline
\textbf{Model} & \textbf{Context Length} & \textbf{Max Output Tokens} & \textbf{Training Data Cutoff} \\
\hline
gpt-4-turbo         & 128,000 tokens  & 4,096 tokens   & Up to Dec 2023 \\
\hline
gpt-4o-mini         & 128,000 tokens  & 16,384 tokens  & Up to Oct 2023 \\
\hline
gpt-3.5-turbo-0125  & 16,385 tokens   & 4,096 tokens   & Up to Sep 2021 \\
\hline
\end{tabular}
\label{tab:model_properties}
\end{table}

The game logic simulates a 20-round word association task, where each model generates a word based on prior exchanges while adhering to predefined constraints.

\paragraph{Model Interaction and Prompting.}  
Each game begins with a system message explaining the rules to both models. At every round, each model receives a prompt that includes:
\begin{itemize}
    \item A reminder that the goal is to produce the same word.
    \item A list of previously used words (which cannot be repeated).
    \item The last word provided by the opponent.
\end{itemize}
The model then generates a single-word response, ensuring that it does not exceed 20 tokens. To encourage diverse word choices and prevent overly deterministic responses, we set the temperature parameter to $1.2$. The game continues iteratively until either:
\begin{itemize}
    \item Both models generate the same word (win condition).
    \item A model repeats a word from previous rounds (loss due to repetition).
    \item A model produces a non-existent word (loss due to invalid input).
    \item The game reaches the 20-round limit without convergence (loss due to non-convergence).
\end{itemize}

\paragraph{Word Validation.}  
To ensure that models only use valid words, we query Wiktionary's API to verify the existence of each word. If a word is not found, the game terminates with a loss due to invalid input.

\paragraph{Dataset.}  
The interaction history for each game is dynamically updated, preserving the sequence of exchanges. The system maintains:
\begin{itemize}
    \item A record of all words exchanged during the game.
    \item The turn-by-turn progression of each model’s responses.
    \item The final game outcome (win, loss due to repetition, loss due to invalid word, or loss due to non-convergence).
    \item The models involved.
\end{itemize}

This dataset serves as the foundation for further analysis, including semantic similarity assessments and strategy evaluations.

\section{Results} \label{results}

\subsection{Model performances}

Table \ref{tab:model_pair_success_rate} summarizes the success rates and average rounds for various model pairings during word association tasks. These results underscore the importance of both model sophistication and compatibility in determining performance.

\begin{table}[ht]
\centering
\caption{Success Rate and Average Rounds for Model Pairs (over 20 trials for same model pairs and over 40 trials when models are differents}
\begin{tabular}{|l|l|l|}
\hline
\textbf{Model Pair} & \textbf{Success Rate} & \textbf{Average Rounds} \\
\hline
(gpt-4-turbo, gpt-4-turbo)               & 100\%  & 3.35 \\
\hline
(gpt-4o-mini, gpt-4o-mini)               & 90\%   & 6.06 \\
\hline
(gpt-4-turbo, gpt-4o-mini)               & 82.5\% & 7.03 \\
\hline
(gpt-3.5-turbo-0125, gpt-3.5-turbo-0125) & 75\%   & 4.6 \\
\hline
(gpt-3.5-turbo-0125, gpt-4o-mini)        & 60\%   & 5.125 \\
\hline
(gpt-3.5-turbo-0125, gpt-4-turbo)        & 50\%   & 4.65 \\
\hline
\end{tabular}
\label{tab:model_pair_success_rate}
\end{table}

The analysis reveals that homogeneous pairings of more advanced models achieve the highest success rates and require fewer rounds to complete the tasks. This trend indicates that advancements in model capabilities directly correlate with improved performance in complex associative tasks. Notably, the pairing of \textit{gpt-4-turbo} with itself achieved a perfect success rate with the least number of rounds, emphasizing the benefits of using the latest AI technologies.

Conversely, pairings involving older or less advanced models, such as \textit{gpt-3.5-turbo-0125}, show decreased success rates and require more rounds to reach the same outcomes, suggesting a potential trade-off between cost and efficiency in deployment scenarios.

\subsection{Embedding Distance Calculation and Visualization}

In order to provide deeper insights into the results, we visualize the progression of the game and how closely the models' word associations align.

To analyze the semantic differences between the words used by both models, we generated embeddings for the word lists, which allow us to measure the semantic distance between the words at each round of the game.

To obtain the word embeddings, we utilized OpenAI's API, specifically the embedding endpoint, which provides high-quality vector representations of text. These embeddings capture semantic information by encoding words into numerical vectors based on the contextual meanings learned by the model. The use of OpenAI's embeddings ensures that the semantic similarities and differences among words are accurately reflected in the vector space.

We then use the Euclidean distance formula to compute the distance between the embeddings generated for both models' words across each round. This distance serves as a proxy for how semantically similar the words were between both participants at each step of the game.

\begin{figure}[ht]
  \centering
  \includegraphics[width=0.8\linewidth]{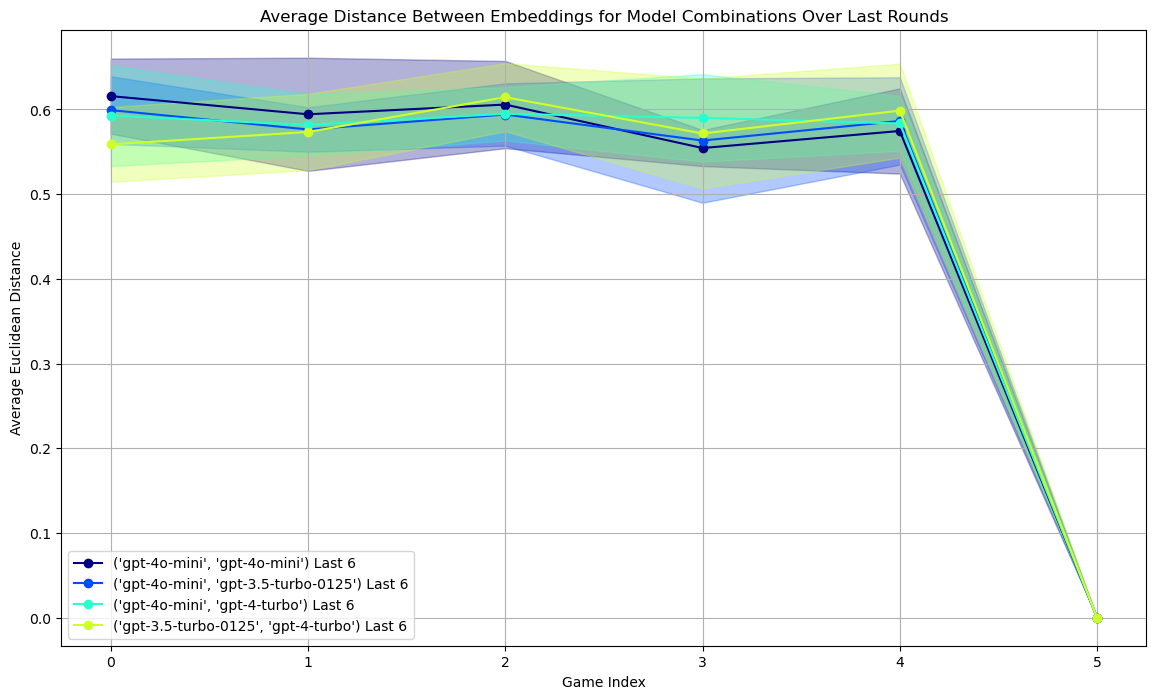}
  \caption{Average Distance Between Embeddings for Model Combinations Over Last Rounds}
  \label{fig:euclidean_distance}
\end{figure}

Figure \ref{fig:euclidean_distance} displays how the semantic similarity changes as the game progresses, indicating whether the models' word choices are converging or diverging.

\subsection{Characterization of the Models' Strategies}

This subsection explores the underlying strategies employed by the LLMs during the word association games. Specifically, we analyze whether the models tend to select words that are semantically closer to the average embedding of the two previous words or if they opt for words closer to the previous word from the opposing model. This analysis provides insights into the decision-making process and strategic behavior exhibited by the models.

For LLMs, this evaluation helps discern whether the associations generated during gameplay are rooted in genuine semantic relationships or if they merely reflect other statistical correlations learned from training data like phonetics or co-usage frequencies.

Our examination involves calculating the Euclidean distances from each model's word choices to two critical reference points:
\begin{itemize}
    \item The embedding of the previous word from the opposing model.
    \item The average embedding of the two last words (one from each model).
\end{itemize}

Models that consistently select words closer to the previous word of their opponent might be employing a \textit{mirroring strategy}. This approach suggests a reactive behavior where the model potentially leverages the context set by the opponent's last word to ensure coherence and continuity in word choice. This behavior might be indicative of models designed to enhance dialogue flow and maintain topic relevance.

Alternatively, selecting words closer to the average embedding of the last two words indicates a \textit{balancing strategy}. This method reflects a more independent and possibly creative approach to word selection, aiming to blend both participants' inputs into a new direction that does not strictly follow the immediate cue but considers the broader context established by both models.

\begin{figure}[ht]
  \centering
  \includegraphics[width=0.8\linewidth]{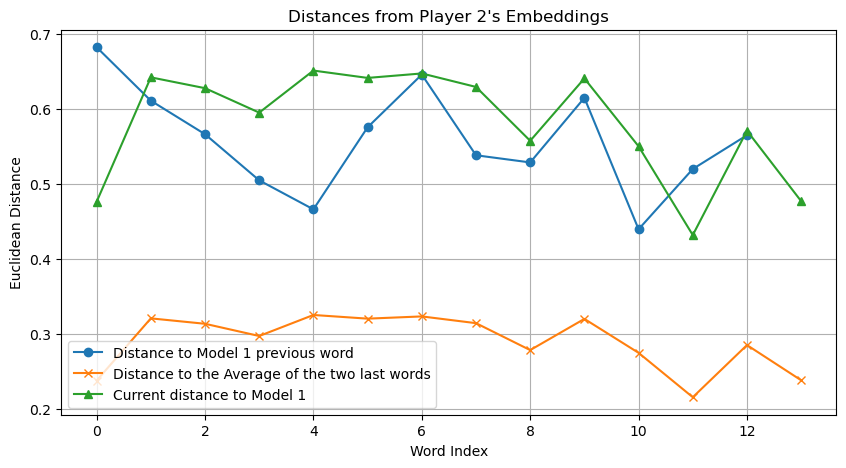}
  \caption{Comparative analysis of model (here gpt-4o-mini playing against himself) preferred strategies showing preferences for word selection relative to the previous word and the average of the last two words}
  \label{fig:strategy_comparison}
\end{figure}

Further, these strategies might also reveal underlying efficiencies or creative capacities inherent in the models. By examining these tendencies, developers can better understand model strengths and weaknesses, aiding in the optimization of AI systems for specific linguistic tasks.

To empirically assess the strategies employed by the models, we implemented a computational analysis using word embeddings and Euclidean distances. For each game where a model won, we calculated two key metrics:

\begin{itemize}
\item \textbf{Mean Distance to Previous Word}: The average Euclidean distance between the embedding of the model’s selected word and the embedding of the opponent’s last word.
\item \textbf{Mean Distance to Average Embedding}: The average Euclidean distance between the embedding of the model’s selected word and the average embedding of the last two words (one from each model).
\end{itemize}

By comparing these two distances, we can infer the predominant behavior of each model. A smaller mean distance to the previous word suggests a \textit{mirroring strategy}, whereas a smaller mean distance to the average embedding indicates a \textit{balancing strategy}.

To perform this analysis, we utilized the following computational approach:

\begin{enumerate}
\item For each model, we filtered the dataset to remove the games where a word and where a word that doesn't exist was outputted.
\item We extracted the embeddings of the past words for both the model and its opponent.
\item We calculated the Euclidean distances between the model’s current word embedding and:
\begin{itemize}
\item The embedding of the opponent’s previous word.
\item The average embedding of the last two words.
\end{itemize}
\end{enumerate}

Fig~\ref{fig:strategy_comparison} shows the evolution of those metrics across an instance of the game. 

We computed the mean of these distances across all relevant games for each model. By comparing the mean distances, we determined the predominant behavior for each model. Results for the three models: \texttt{gpt-4-turbo}, \texttt{gpt-4o-mini}, and \texttt{gpt-3.5-turbo-0125} are summarized in Table~\ref{tab:strategy_results}. All three models exhibit a smaller mean distance to the average embedding compared to the distance to the previous word. This indicates that the models predominantly employ a \textit{balancing strategy} during gameplay. The balancing strategy reflects an approach where the model integrates the context from both its own last word and the opponent’s last word to generate a new word that advances the game in a coherent and creative manner.

\begin{table}[ht]
\centering
\caption{Mean distances and predominant strategies for each model (we count the games were they are playing against themselves twice)}
\begin{tabular}{|l|c|c|c|c|}
\hline
\textbf{Model} & \textbf{Dist. to Previous} & \textbf{Dist. to Average} & \textbf{ Samples} & \textbf{Main Strategy} \\
\hline
gpt-4-turbo        & 0.513 ± 0.040 & 0.449 ± 0.028 & 77 & Balancing \\
\hline
gpt-4o-mini        & 0.552 ± 0.040 & 0.477 ± 0.031 & 79 & Balancing \\
\hline
gpt-3.5-turbo-0125 & 0.539 ± 0.033 & 0.470 ± 0.026 & 60 & Balancing \\
\hline
\end{tabular}
\label{tab:strategy_results}
\end{table}

These findings suggest that the models are inclined towards maintaining a balanced semantic progression rather than simply reacting to the opponent’s immediate input. This behavior can be advantageous in tasks that require innovation and the synthesis of ideas, such as brainstorming sessions or collaborative writing.

In summary,\texttt{gpt-4-turbo}, \texttt{gpt-4o-mini}, and \texttt{gpt-3.5-turbo-0125} predominantly employ a balancing strategy to solve the Word Synchronization Challenge.

\subsection{Dynamics evaluations}

Finally, the visualization of word embedding trajectories offers crucial insights into how two models adjust their language choices over time to align with each other’s cognitive models. By tracking these embeddings, we can observe the dynamic process of synchronization, which is fundamental to effective communication. In collaborative interactions, the convergence of word embeddings signals the achievement of a shared understanding, where both models coevolve toward the articulation of the same or semantically similar words. This highlights successful coordination and cooperation, essential elements for real-world Human-Computer Interaction (HCI). The ability to dynamically align on language use is not just a measure of technical capability but also a proxy for social intelligence, as chatbots that can synchronize with human partners demonstrate greater communicative competence. This alignment process is particularly significant in HCI, where chatbots must continually adapt to human behaviors, ensuring that their responses are perceived as natural and contextually appropriate.

In particular, we can check if the words chosen by the models are within certain manifolds which give further understanding into their behaviors.

The word embeddings were merged and reduced to 3 dimensions using PCA. This allowed for clear visualization of how the words evolve and converge over time in the word association game. We visualized the embeddings in a 3D scatter plot. The scatter plot represents the words trajectories in the embedding space of both models over time. If the final words match, the last word is highlighted with a special marker (a star or diamond). This visualization offers a clear understanding of how each models’s word association trajectory unfolds and whether or not the game achieves synchronization.

\subsubsection{Example of an unsuccessful case}

\begin{figure*}[t]
  \centering
  \begin{subfigure}{0.45\linewidth}
    \centering
    \includegraphics[width=\linewidth]{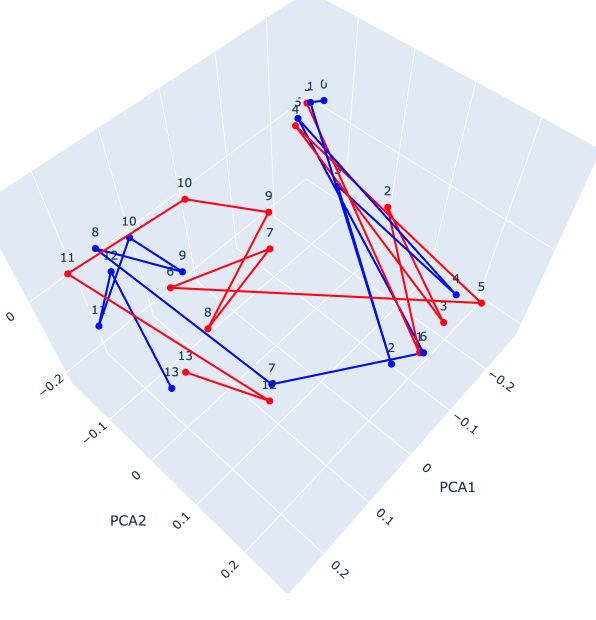}
    \caption{View from above of the projection}
    \label{fig:lost_view_above}
  \end{subfigure}%
  \hfill
  \begin{subfigure}{0.45\linewidth}
    \centering
    \includegraphics[width=\linewidth]{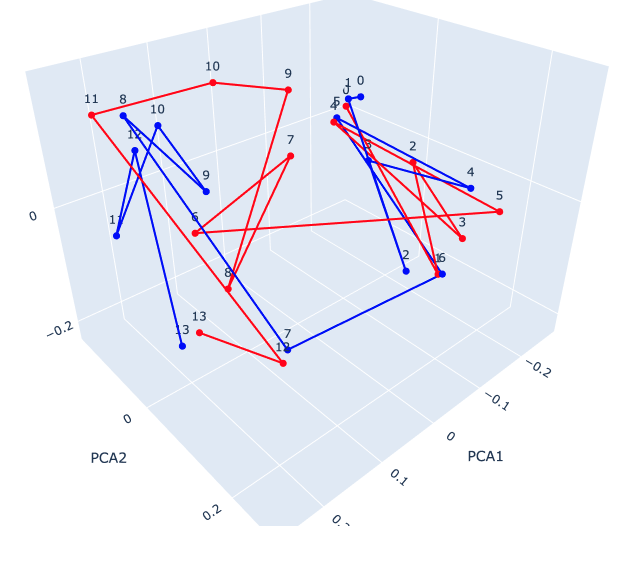}
    \caption{View from the side of the projection}
    \label{fig:lost_view_side}
  \end{subfigure}
  \caption{Two different views of the projection of the embedding of one game between two instances of GPT-4o-mini}
  \label{fig:unsuccessful_manifolds}
\end{figure*}

Figure \ref{fig:unsuccessful_manifolds} shows that the models tried to converge in 3 manifolds in space, in particular we see in the first utterance jumps between the two first manifolds, indicating that the models where trying to get closer to each other by picking a word that was just close semantically to the previous word of the other model. This explanation is corroborated when we take a closer look at the words themselves. 

\begin{table}[ht]
\centering
\caption{Sequence of words exchanged involving GPT-4o-mini during a game, categorized by semantic themes and color-coded accordingly}
\begin{tabular}{|c|c|c|c|c|c|c|c|}
\hline
\textbf{Model} & 1 & 2 & 3 & 4 & 5 & 6 & 7 \\
\hline
\textbf{GPT-4o-mini} & \textcolor{darkgreen}{Banana}    & \textcolor{darkgreen}{Mango}   & \textcolor{blue}{Sky}      & \textcolor{darkgreen}{Orange}  & \textcolor{blue}{River}   & \textcolor{darkgreen}{Lemon}   & \textcolor{blue}{Stream} \\
\hline
\textbf{GPT-4o-mini} & \textcolor{darkgreen}{Pineapple} & \textcolor{blue}{Cloud}        & \textcolor{darkgreen}{Apple}   & \textcolor{blue}{Ocean}   & \textcolor{darkgreen}{Citrus}  & \textcolor{blue}{Lake}    & \textcolor{red}{Sunshine} \\
\hline\hline
 & 8 & 9 & 10 & 11 & 12 & 13 & 14 \\
\hline
\textbf{GPT-4o-mini} & \textcolor{red}{Light}     & \textcolor{red}{Dawn}     & \textcolor{red}{Twilight} & \textcolor{red}{Sunrise}   & \textcolor{red}{Daylight} & \textcolor{red}{Dusk}     & \textcolor{red}{Illumination} \\
\hline
\textbf{GPT-4o-mini} & \textcolor{red}{Horizon}   & \textcolor{red}{Glow}     & \textcolor{red}{Morning}  & \textcolor{red}{Evening}   & \textcolor{red}{Daybreak} & \textcolor{red}{Brightness} & \textcolor{red}{Radiance} \\
\hline
\end{tabular}
\label{tab:first_table}
\end{table}

 This analysis underscores the models' attempt to achieve semantic proximity through strategic word selection.

\begin{itemize}
    \item \textbf{Fruits (Green)}: The terms \textit{Banana}, \textit{Mango}, \textit{Orange}, \textit{Lemon}, \textit{Pineapple}, \textit{Apple}, and \textit{Citrus} are rooted in tangible, everyday items. The models frequently selected these terms, which may indicate a preference for concrete and familiar concepts, potentially facilitating easier semantic clustering.

    \item \textbf{Sky and Water Elements (Blue)}: Words such as \textit{Sky}, \textit{River}, \textit{Stream}, \textit{Cloud}, \textit{Ocean}, and \textit{Lake} fall into this category. These selections point towards a thematic orientation around natural scenery and elements.

    \item \textbf{Light-Related Phenomena (Red)}: This group includes \textit{Light}, \textit{Dawn}, \textit{Twilight}, \textit{Sunrise}, \textit{Daylight}, \textit{Dusk}, \textit{Illumination}, \textit{Sunshine}, \textit{Glow}, \textit{Morning}, \textit{Evening}, \textit{Daybreak}, \textit{Brightness}, and \textit{Radiance}, which touch upon various aspects of light.
\end{itemize}

Notably, the initial utterances showed transitions between the first two manifolds (fruits and sky/water), revealing that the models employed a local mirroring strategy. By selecting words that were semantically close to the previous word from the other model, they attempted to bridge semantic gaps and maintain coherence in the conversation. For instance, the transition from \textit{Mango} to \textit{Sky} could be conceptually linked by color or as a juxtaposition of natural elements, which, while categorically distinct, share a relational context.

These transitions reflect a sophisticated modeling approach, aiming to enhance narrative fluidity by mimicking human-like thought processes and conversational transitions. The selection patterns suggest that the underlying algorithms are designed not only for coherent word choice but also for enriching the dialogue or text generation with semantic depth and relational intricacies that mirror human discourse.

\subsubsection{Example of a successful case}

\begin{figure*}[t]
  \centering
  \begin{subfigure}{0.45\linewidth}
    \centering
    \includegraphics[width=\linewidth]{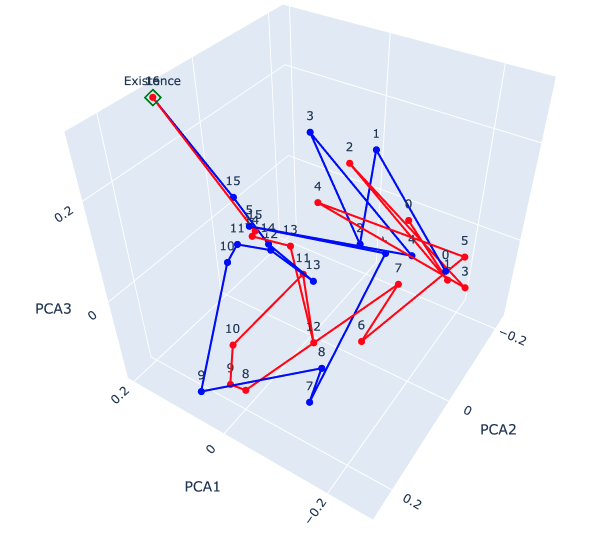}
    \caption{View from above of the projection}
    \label{fig:win_view_above}
  \end{subfigure}%
  \hfill
  \begin{subfigure}{0.45\linewidth}
    \centering
    \includegraphics[width=\linewidth]{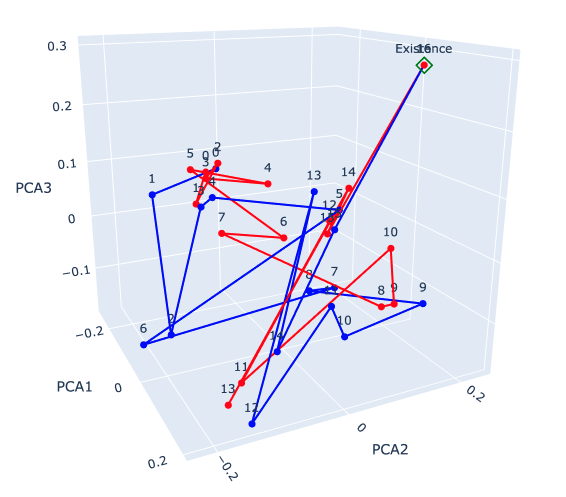}
    \caption{View from the side of the projection}
    \label{fig:win_view_side}
  \end{subfigure}
  \caption{3D plots with two different views of the projection of the embedding for a won game between GPT-4-turbo and GPT-4o-mini}
  \label{fig:successful_manifolds}
\end{figure*}

In a remarkable instance of convergence, the two model outputs aligned showcasing an advanced degree of thematic and semantic synchronization. Both sequences share a deep focus on ecological and environmental themes, with the first placing a stronger emphasis on conservation, while the second delves into biological diversity. The semantic progression in each is meticulously structured, showcasing the models' capabilities in maintaining thematic coherence. The convergence on the term \textit{Existence} underlines a sophisticated conceptual synthesis, reflecting the models' capacity for deep semantic integration.

\begin{table}[ht]
\centering
\caption{Sequence of words exchanged between GPT-4-turbo and GPT-4o-mini models during a successful game, categorized by semantic themes and color-coded accordingly}
\begin{tabular}{|l|l|l|l|l|l|l|l|}
\hline
\textbf{Model} & 1 & 2 & 3 & 4 & 5 & 6 & 7 \\
\hline
\textbf{GPT-4-turbo} & \textcolor{darkgreen}{Elephant} & \textcolor{violet}{Yellow}    & \textcolor{darkgreen}{Australia} & \textcolor{violet}{Color}     & \textcolor{darkgreen}{Safari}  & \textcolor{violet}{Sunlight} & \textcolor{darkgreen}{Africa} \\
\hline
\textbf{GPT-4o-mini} & Banana                          & \textcolor{darkgreen}{Kangaroo} & \textcolor{violet}{Crimson}    & \textcolor{darkgreen}{Giraffe} & \textcolor{violet}{Shade}     & \textcolor{darkgreen}{Zebra}  & \textcolor{blue}{Jungle} \\
\hline\hline
 & 8 & 9 & 10 & 11 & 12 & 13 & 14 \\
\hline
\textbf{GPT-4-turbo} & \textcolor{blue}{Rainforest} & \textcolor{blue}{Grassland} & \textcolor{blue}{Ecosystem} & \textcolor{blue}{Environment} & \textcolor{darkyellow}{Nature} & \textcolor{red}{Earth}   & \textcolor{darkyellow}{Animals} \\
\hline
\textbf{GPT-4o-mini} & \textcolor{blue}{Savanna}    & \textcolor{blue}{Ecozone}   & \textcolor{blue}{Biomes}    & \textcolor{blue}{Habitat}     & \textcolor{blue}{Terrain}  & \textcolor{darkyellow}{Wildlife} & \textcolor{red}{Planet} \\
\hline
\multicolumn{8}{c}{} \\[-2.5ex] 
\cline{0-3}
 & 15 & 16 & 17 \\
\cline{0-3}
\textbf{GPT-4-turbo} & \textcolor{red}{World}     & \textcolor{red}{Life}     & \textcolor{red}{Existence} \\
\cline{0-3}
\textbf{GPT-4o-mini} & \textcolor{darkyellow}{Creatures} & \textcolor{red}{Universe} & \textcolor{red}{Existence} \\
\cline{0-3}
\end{tabular}
\label{tab:second_table}
\end{table}

The analysis of the model outputs reveals the presence of distinct semantic manifolds, each representing a cluster of interrelated themes. These manifolds demonstrate the models' capability to categorize and conceptualize complex information through semantic proximity. Below, we outline these manifolds, which we have identified and defined based on our interpretation of the data, describing their core themes and semantic implications.

\begin{itemize}
    \item \textbf{Wildlife and Animals (Green)}: Encompassing terms such as \textit{Elephant}, \textit{Australia}, \textit{Safari}, \textit{Africa}, \textit{Kangaroo}, \textit{Giraffe}, and \textit{Zebra}, this manifold highlights the models' focus on biological and zoological aspects. 

    \item \textbf{Physical Attributes and Color (Violet)}: Comprising \textit{Yellow}, \textit{Color}, \textit{Crimson}, and \textit{Shade}, this manifold addresses the sensory and perceptual attributes of objects, focusing on visual elements.

    \item \textbf{Natural Landscapes and Locations (Blue)}: This theme includes \textit{Jungle}, \textit{Rainforest}, \textit{Grassland}, \textit{Ecosystem}, \textit{Environment}, \textit{Savanna}, \textit{Ecozone}, \textit{Biomes}, \textit{Habitat}, and \textit{Terrain}. It reflects a transition toward a manifold in  complex environmental interactions and natural formations.

    \item \textbf{General Nature and Wildlife (Yellow)}: With terms like \textit{Nature}, \textit{Animals}, \textit{Wildlife}, and \textit{Creatures}, this grouping broadens the scope to general biological and natural concepts, offering a macroscopic view of life and natural entities. It anchors the more detailed observations found in other manifolds, providing a foundational context for understanding life's diversity.

    \item \textbf{Universal and Abstract Concepts (Red)}: This category, including \textit{Earth}, \textit{Planet}, \textit{World}, \textit{Life}, \textit{Existence}, and \textit{Universe}, enters the realm of the existential and metaphysical. It illustrates the models' profound capacity to engage with abstract concepts that are broader compared to the previous manifolds.
\end{itemize}

Compared to the previous unsuccessful example, the models in this successful case exhibited a more nuanced approach in their word selection strategies, effectively engaging in both mirroring and balancing strategies. This dual engagement allowed them to reflect the opponent’s immediate semantic cues while also integrating the broader context of the conversation. By alternating between these strategies, the models navigated through complex semantic manifolds, progressively aligning their thematic focus.

The initial exchanges show the models selecting words within the same semantic manifolds, such as wildlife and animals, and physical attributes and colors. This mirroring strategy facilitated immediate semantic coherence. As the conversation progressed, they shifted towards a balancing strategy, considering the average embedding of the previous words to introduce new but contextually relevant themes, such as natural landscapes and universal concepts.

In contrast to the unsuccessful case, where the models remained confined within limited semantic manifolds and failed to effectively bridge between them, the successful interaction showcases the importance of strategic flexibility. This might be due to the fact that the most capable model, GPT-4-turbo, was this time involved.

\section{Discussion} \label{discussion}

\subsection{Conclusions} \label{further}

The Word Synchronization Challenge is designed as a flexible game that accommodates various player configurations. It can be played between two human participants, two Large Language Models (LLMs), or a hybrid pairing of one human and one LLM. This adaptability allows for different types of interactions, offering opportunities to study communication dynamics in diverse settings. When played between two humans, the game highlights natural language understanding and the intuitive connections that people can establish. When played between two LLMs, it examines how AI systems generate and align language based on learned patterns. The mixed mode, where a human pairs with an LLM, provides insight into the extent to which LLMs can synchronize with human reasoning, which may inform future research in human-AI collaboration.

\begin{figure}[ht]
  \centering
  \includegraphics[width=0.35\linewidth]{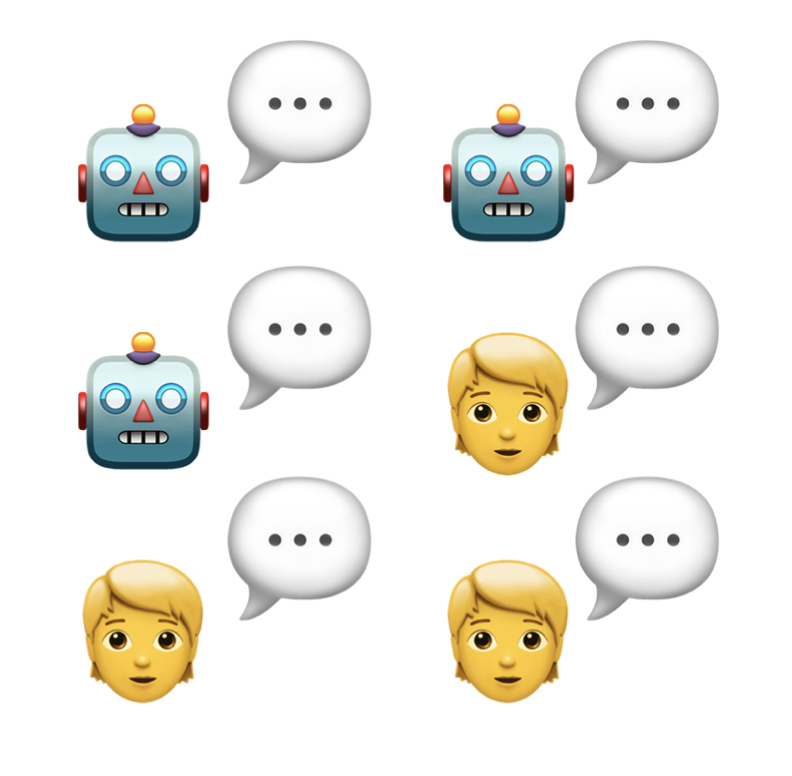}
  \caption{The game can be played between two LLMs, two humans, or between an LLM and a human, which allows for a promising development of this benchmark}
  \label{fig:players}
\end{figure}

While the game mechanics are straightforward, achieving consistent synchronization requires representational and inferential abilities that have traditionally been associated with human cognition. This characteristic makes the Word Synchronization Challenge a potentially useful tool for studying aspects of interpersonal and computational communication, such as theory of mind, social perception, and adaptive coordination. However, the extent to which the game effectively captures these cognitive processes remains an open question, as performance may be influenced by various factors, including prior knowledge and linguistic biases.

When considering LLM-LLM games, a key consideration in the game’s setup is the method of word initialization. Non-random word initialization, where starting words are selected based on frequency, semantic relatedness, or other heuristic criteria, may introduce biases that affect synchronization outcomes. We observed that LLM-generated starting words tend to be remarkably simple, often consisting of high-frequency, common words, which may reduce the challenge of synchronization and favor strategies based on statistical co-occurrence rather than genuine mutual understanding. This could lead to inflated performance estimates, particularly for LLMs that rely on learned language patterns rather than active coordination. Conversely, random initialization may introduce greater unpredictability, making the task more reflective of real-time communicative adaptation but also increasing variability in results.

Overall, while the Word Synchronization Challenge offers a structured way to explore synchronization and mutual understanding, further investigation is needed to assess its reliability as a benchmark for complex cognitive and social skills. Its effectiveness as a tool for evaluating human-like communication and coordination depends not only on the nature of the participants but also on the methodological choices underpinning its implementation.

\subsection{Future work} \label{futur_work}

We created an interface to collect data from humans.\footnote{The interface is available online at \url{https://word-sync.games/}}.

\begin{figure}[ht]
  \centering
  \includegraphics[width=0.8\linewidth]{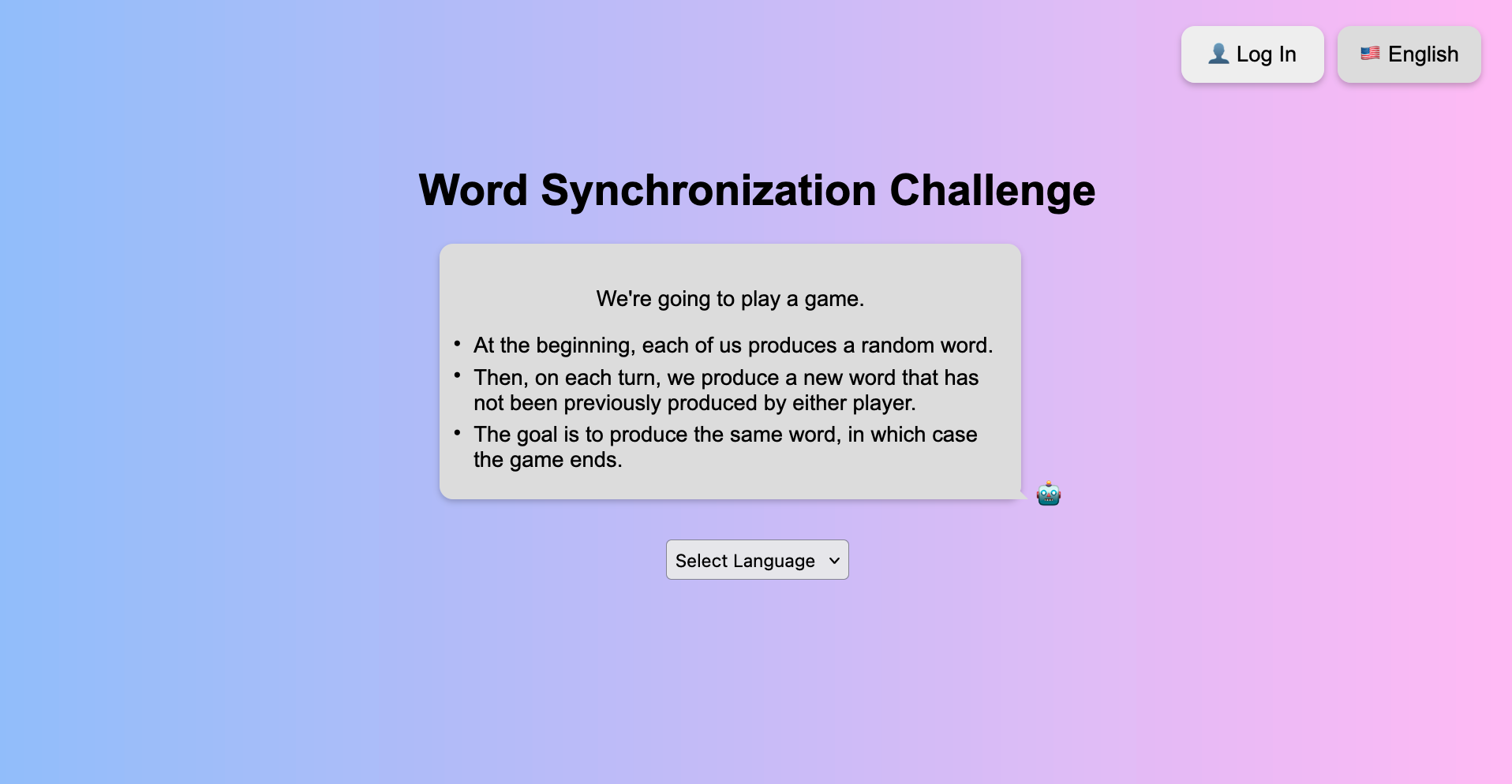}
  \caption{Screenshot of the interface built to collect human data}
  \label{fig:interface}
\end{figure}

Currently, the model used in this application can only be selected from a variety of pre-trained small models available on Hugging Face. However, we plan to extend this functionality in the near future by integrating the application with OpenAI’s API. his extension will grant access to more advanced and continuously updated models, improving the app’s ability to handle complex reasoning and dynamic interactions.

Beyond model integration, we aim to leverage the collected human interaction data to explore deeper cognitive mechanisms underlying word association and adaptation processes. The reasoning patterns behind user word choices serve as valuable signals for fine-tuning models, potentially improving their ability to mimic human reasoning and align more naturally with human communicative strategies. Additionally, the trajectory of word selections across multiple rounds offers insights into adaptive behaviors, shedding light on how users adjust their strategies over time in response to LLM interactions.

In future iterations, this app will be employed to investigate both human-human and human-LLM interactions, expanding its utility as a tool for studying collaborative language games. By systematically analyzing user behavior in these settings, we aim to uncover key mechanisms of alignment and divergence between human and machine communication. Comparing LLM-generated associations with human associative patterns will provide crucial insights into the internal representations of meaning within these models, informing broader research on human cognition and AI alignment.

Ultimately, this work contributes to a deeper understanding of the bidirectional nature of human-AI adaptation. As language models become increasingly integrated into real-world applications, it is essential to quantify and interpret the ways in which they shape human communication. By analyzing the interplay between human and machine word selection, we can refine AI systems to better support naturalistic, collaborative, and adaptive human interactions while preserving the diversity and nuance of human cognition.

\begin{credits}
\subsubsection{\ackname} This project has received funding from the European Union’s Horizon 2020 research and innovation program under the Marie Skłodowska-Curie grant agreement No 860949.
\end{credits}

\bibliographystyle{splncs04}
\bibliography{main}

\end{document}